\documentclass[twocolumn]{revtex4}
\usepackage{graphicx}
\usepackage{epstopdf}
\usepackage{longtable}

\begin{document}

\title{Heisenberg, Uncertainty, and the Scanning Tunneling Microscope.}

\author{Werner A. Hofer}
\affiliation{Department of Physics, University of Liverpool\\
L69 3BX Liverpool, United Kingdom}

\begin{abstract}
We show by a statistical analysis of high-resolution scanning tunneling microscopy (STM) experiments, that the interpretation of the density of electron charge as a statistical quantity leads to a conflict with the Heisenberg uncertainty principle. Given the precision in these experiments we find that the uncertainty principle would be violated by close to two orders of magnitude, if this interpretation were correct. We are thus forced to conclude that the density of electron charge is a physically real, i.e. in principle precisely measurable quantity.
\end{abstract}

keywords: scanning tunneling microscope, electron charge, density functional theory, uncertainty relations

\maketitle

Today, STMs have reached a level of precision which is quite astonishing. While it was barely possible to resolve the positions of single atoms in early experiments \cite{BIN82,BIN87}, it is now routine not only to resolve the atomic positions but also e.g. the standing wave pattern of surface state electrons \cite{HEL94} on metal surfaces, and even very subtle effects like inelastic excitations, Kondo resonances, or surface charging \cite{GAW08,NEE07,HAR08}. What the STM measures, in these experiments, is the current between a surface and a very sharp probe tip.

The current itself is proportional to the density of electron charge at the surface \cite{HOF03}. While one may dispute this claim for some special cases, and while it can be shown for specific situations that an explicit simulation of the scattering process does improve the agreement between experiment and theory (see, for example \cite{DENG06}), in measurements on metal surfaces the bias range is so low and the dominance of single electron-states at the tip so high, that the Tersoff-Hamann approximation \cite{TH85}, which assumes tunneling into a single tip-state with radial symmetry, is a very good approximation.  Then the map of tunneling currents at a surface is, but for a constant, equal to the map of electron charge densities at the same surface. A standard deviation of the density of charge due to the uncertainty of position and momentum can thus be mapped identically onto a standard deviation of the tunneling current, which can immediately be compared to experimental results.

From the viewpoint of a physicist of the golden age of quantum mechanics, say the time around 1925, the precision of STM measurements, where the vertical resolution of the best instruments is close to fifty times smaller (about 0.05 pm \cite{GAW08}) than the Compton wavelength of an electron (or about 2.4 pm), must seem like magic. In Figure 1 we show a quantum corral of 51 silver atoms on an Ag(111) surface measured by Rieder et al. \cite{RIE04}. In some images the instrument was able to resolve not only the standing wave pattern of surface state electrons, but also the modulation due to the positions of single surface atoms. The measurement is quite spectacular, and it raises some very fundamental problems in quantum mechanics, which we shall explore in the following. It is irrelevant for this analysis, whether the remaining experimental imprecision is due to microscopic or macroscopic processes: it is the level of precision in these experiments itself, which is problematic.

In density functional theory (DFT) a many-electron system is comprehensively described by the density of electron charge \cite{HK64,KS65}. However, the density itself, within the framework of second quantization, is thought to be a statistical quantity \cite{BORN26}. In principle, this statement can be tested by a statistical analysis of high-resolution STM measurements including the uncertainty relations \cite{HEI27}. The reason that such an analysis at this point is necessary is the following:

In a recent paper \cite{HOF11} it was found that an extended model of electrons fully recovers the formulation of quantum mechanics in terms of Schr\"odinger's equation \cite{SCHR26}. This leads to a very fundamental issue in quantum mechanics. Originally, it was found by Born, Jordan and Schr\"odinger that the formulation of the theory in terms of the Schr\"odinger equation is equivalent to the formulation in terms of operators and matrices, as proposed by Heisenberg \cite{HEI25}, implying the uncertainty relations. However, this equivalence depends on the interpretation of the electron as a point-particle. It is not a priori valid for an extended electron. For extended electrons the density of charge will be a physical property of a microsystem. The question then arises, whether the uncertainty relations are still meaningful in this context. This question can in principle be answered by an analysis of measurements of electron charge distributions. It should be noted that the
interpretation of the electron wave as a physical wave makes the usual, purely mathematical, derivation of the Uncertainty
Relations via Fourier transforms and a Gaussian wave packet untenable. If the electron is a real wave, then it does possess
a physical momentum and it is essentially a plane wave confined to some region in space. In this case a composition of single electrons in terms of a Fourier series of partial waves with different momenta, even though mathematically possible, is physically not meaningful. The Uncertainty Relations are thus a statement about the physical properties of systems at the atomic scale \cite{MARG68}. If it turns out that even in our most precise measurements the interpretation of the density of electron charge as a statistical quantity is still possible, then the uncertainty relations have a meaning also in this context. However, if it turns out that standard statistical quantities, depending on the uncertainty of location and momentum, will be much smaller in the actual experiments than allowed for by a theoretical analysis, then the uncertainty relations, and by implication operator mechanics, cannot be considered an equivalent description. In this case, it should also be ultimately possible to measure any physical property of a microsystem, i.e. an atomic-scale system, with in principle infinite precision.

The statistical measure we shall use for our analysis is the standard deviation $\sigma$. The standard deviation for a measurement of position $x$, given $N$ measured values $x_i$, is defined as:
\begin{equation}
\sigma_x = \sqrt{\frac{1}{N} \sum_{i=1,N} \left(x_i - \langle x \rangle\right)^2} \qquad \langle x \rangle = \frac{1}{N} \sum_{i=1,N} x_i
\end{equation}

where $\langle x \rangle$ denotes the statistical average. We shall analyze the variation of the density of electron charge at a metal surface, and compare the local contrast in the experiments, which is given by the change of the density from one point of measurement to the next, with the achievable contrast on the basis of allowed energy, momentum, and position uncertainty. We shall explicitly treat noble metal (111) surfaces, where free electrons, confined to the surface layer, exist, which can be treated for all practical purposes as free electrons. They should thus comply with the following fundamental relation in quantum mechanics:
\begin{equation}
\Delta p \cdot \Delta x \ge \frac{\hbar}{2}
\end{equation}
The relation between an uncertainty in momentum and an uncertainty in energy is described by:
\begin{equation}
\Delta E = \Delta \left(\frac{p^2}{2 m}\right)
\end{equation}

It is difficult to define the exact uncertainty of momentum of a surface state electron in the experiments. From the energy resolution in today's best measurements of about $\Delta E = 1$meV \cite{HEI04} and using Eq. (2) one could infer an uncertainty of momentum of $\Delta p = p = 1.52\times10^{-26}$ kgms$^{-1}$, and a local uncertainty $\Delta x = 3.1\times10^{-9}$ m, or 3.1 nm. However, to establish that the uncertainty relations fail in this context, it is only necessary to provide an upper limit of the energy and momentum uncertainty. Thus, for an electron at the Fermi level of a Ag(111) surface the maximum available energy is given by the energy difference between the bottom of the parabolic surface state band and the energy at the Fermi level.  This energy is about 80 meV. This corresponds also to the thermal energy at ambient conditions. As the experiments are performed at cryogenic temperatures of 5K, the thermal energy is much lower than the band energy. Hence the maximum uncertainty in the energy of the electron is also about 80 meV. From this value it is possible, via the energy uncertainty and the Heisenberg uncertainty relations (Eqs. (2)), to derive the uncertainty of the x-position: $\Delta x = 3.48 \times 10^{-10}$m  or about 3.5 \AA. This value is about three times larger than the Wigner-Seitz radius (1.06 \AA), which describes the radius of a sphere with the typical electron density found in metals. The uncertainty $\Delta x$ is also larger than the distance between two individual atoms on a silver surface, which is 2.9 \AA. The unit cell, the Wigner-Seitz cell, and the uncertainty are marked in Figure 1c.

\begin{figure}
\centering
\includegraphics[width=\columnwidth]{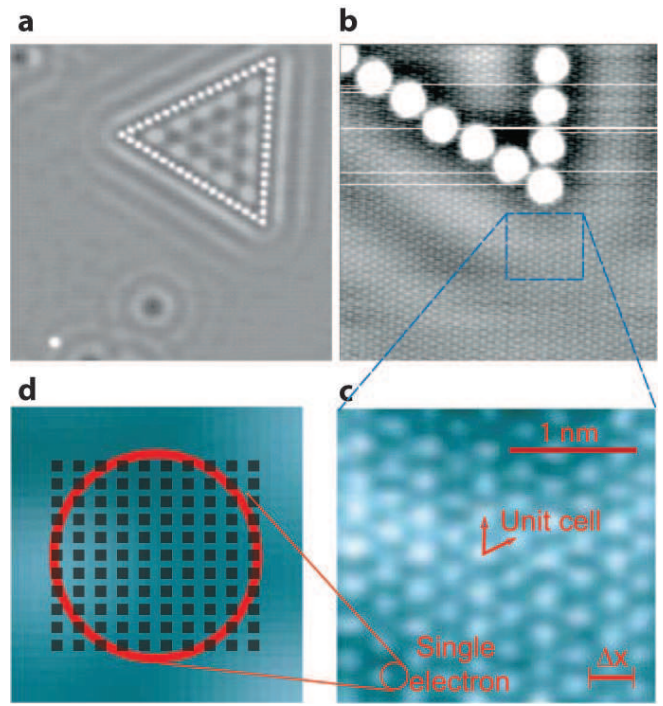}
\caption{(online version in colour): Quantum corral and surface state electron on a Ag(111) surface. a The propagation of the surface state electrons leads to periodic variations of the density and the apparent height. b In high-precision measurements these variations are modulated by the electronic structure of surface atoms. c Small region of the central image blown up and annotated: the unit cell, the area of a single electron, and the local uncertainty are marked. d The Wigner-Seitz cell of a single electron contains at least 88 pixels of measurements, if the lateral resolution is assumed to be 20pm. From Ref. \cite{RIE04}}
\end{figure}

Following the standard interpretation of quantum theory the electron is a point particle. Its appearance at a particular location is a statistical event governed by the Schr\"odinger equation and the probability density, which is equal to the square of the electron's wavefunction, or $|\psi|^2$. The same statement is also correct for alternative interpretations of quantum mechanics, e.g. based either on the pilot-wave theory \cite{BOH52,DEB60}, or a theory of the Dirac electron based on Zitterbewegung \cite{HES10}. In this case the precision of an experiment will depend on the statistical variation of the electron's location and its convolution over time. This, however, will also influence the statistical variation of e.g. the apparent height. If the apparent height can be measured with a precision of less than 0.05pm, as found in experiments \cite{GAW08}, then the statistical probability of an electron to be found at a particular location can only have a certain standard deviation $\sigma$. Given the local uncertainty derived from the uncertainty relations, this deviation can, in principle, be linked either to a minimum number of events within the measurement interval of an STM, which is in the range of milliseconds, or to the error in the vertical distance measurement combined with the lateral resolution in the experiments. Note that the argument differs substantially from
previous analysis of STM measurements, which only considered the vertical precision: given the time resolution of the instrument
and a tunneling current of 1nA about 10$^7$ electrons cross the vacuum barrier within the measurement resolution of the instrument. This, alone, will not be sufficient to establish the statistical facts. Only the {\it lateral resolution} in
combination with the {\it vertical precision} will be sufficient to analyze the statistics in these experiments comprehensively.

We have thus two possible ways in which the experimental values can be compared to the standard deviation inferred from the uncertainty relations. We may either find, that a threshold number of measurements N$_0$ is necessary to obtain the experimentally observed standard deviation. In this case we can estimate the minimum distance covered by the electron within the time-resolution limit of STM experiments $\Delta t(STM)$ (i.e. milliseconds), and derive a lower limit for the time-resolution of the instrument  without a violation of the uncertainty relations. Such a limit is described by:
\begin{eqnarray}
\Delta L = N_0 dl \quad dl \approx \Delta x \quad v(\Delta E) \Delta t (STM) \ge \Delta L
\nonumber \\ \Delta t (STM) \ge \frac{\Delta L}{\sqrt{2 \Delta E/m}}
\end{eqnarray}

where $\Delta L$ is the total path during N$_0$ measurements, $dl$ is the average path from one measurement to the next, and $\Delta E$ is the energy uncertainty. Or we may find that the calculated standard deviation is actually quite independent of the number of measurements. In this case the value of the standard deviation for an energy uncertainty $\Delta E$ can be directly compared to the maximum standard deviation of the tunneling current given the precision of the instrument. Here, the percentage of error in an STM measurement, calculated from ratio of the error in the distance measurement divided by the apparent height of a feature, can be used, in combination with the lateral resolution of the instrument, to determine the maximum standard deviation. For an error in the experiments of 0.3\%, which is easily obtained in today's most precise experiments, we find then:
\begin{equation}
\frac{\Delta z}{z_0} \le 0.3\mbox{\%} \quad \Rightarrow \quad \Delta x (STM) \ge 3 \sigma (\Delta E)
\end{equation}
where $\Delta x(STM)$ is the lateral resolution of an STM and $\Delta E$ the maximum energy uncertainty. If three times the standard deviation inferred from the energy uncertainty is substantially larger than the lateral experimental resolution $\Delta x(STM)$, then one of the original assumptions going into the analysis of the process must be flawed. We show that this is indeed the case.

In the following we start the discussion by analyzing events at one particular pixel of the grid on the surface and analyze events from the viewpoint of one electron only. Subsequently, we shall determine how the presence of many electrons at the surface could alter the initial findings.

The apparent height of a feature measured by STM is, for a metal surface, in the range of 10 to 200pm. For the sake of the argument, and to avoid being overly restrictive, we assume the vertical resolution to be 0.1pm, and the lateral resolution to be 20pm. In a Wigner-Seitz cell we then have about 88 individual pixels of STM measurements (see Figure 1d). If we assume that the feature height is 30pm, then the experimental precision is 0.3\%. The apparent height at an individual pixel would be affected, however, if at a given moment during the measurement interval the electron were not found at the location of this pixel, but at a neighboring pixel. Thus we may say that the standard deviation of the location measurement is such that the required percentage of measurements must be at this particular location. For a variation of less than 0.3\% three times the standard deviation of the measured results must thus still be smaller than half the distance between the pixels (see Figure 1d), or about 10pm. The standard deviation must thus be smaller than 3.3pm. Given that the uncertainty $\Delta x = 348$pm, every single measurement event, at which the electron should be detected at the position of one pixel at $(x, y) = (0, 0)$, will lead to a statistically distributed measurement in the interval $-348 \le x \le +348$, and $-348 \le y \le 348$. We have simulated such a statistical distribution of events in Figure 2. It can be seen that the locations of the electron are indeed randomly distributed in the interval, as required by the uncertainty relations and quantum mechanics. We find a standard deviation of about 280pm for all sets of data. Increasing the number of measurements does not change the precision, since the precision is limited by the uncertainty relations themselves. The precision can only be increased by an increase of the electron's energy. Since $\sigma$ is about one hundred times the required value, we may say that there is a clear contradiction between the observed precision of the experiments, and the calculated uncertainty for the location of the electron. It is also clear that the underlying assumption, leading to this contradiction, is the statistical nature of the measurement, encoded in the uncertainty relations. This contradiction, moreover, is not the result of a small variation of the values obtained, but a variation of two orders of magnitude.
\begin{figure}
\centering
\includegraphics[width=6cm]{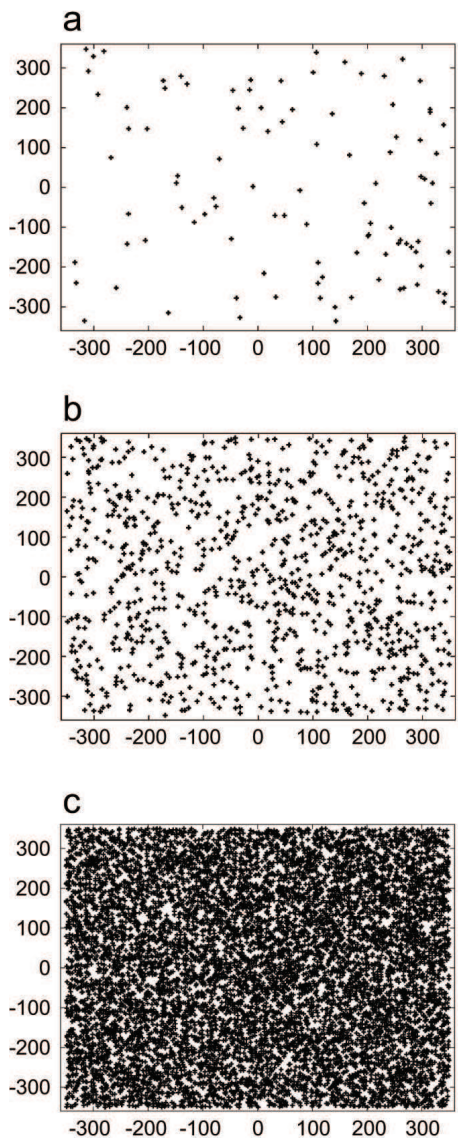}
\caption{Random distribution of the location of a point-like electron on the surface. a 100 events, b 1000 events, c 10000 events. The standard deviation   of the radius for each of the distributions is about 280 pm (random distribution taken from random.org).}
\end{figure}
It is straightforward to calculate the energy necessary to obtain the precision found in the experiments. If $\sigma$ has to be around 3pm, then the necessary uncertainty   will also be about 3pm. This means that the momentum uncertainty has to be around $1.76 \times 10^{-23}$kgms$^{-1}$, and the necessary energy will be more than 1000 eV. For a measurement taken at cryogenic temperatures of 5K, as routinely performed today, the probability of such energy, derived from the Fermi-Dirac distribution, is essentially zero; the precision found in the experiments is thus impossible to explain within the framework of standard quantum theory.

The result would not change substantially if the precision in the experiments were somewhat lower. However, so far we have assumed that the uncertainty leads to a loss of electron detection at a particular point. It is, at least in principle, also possible, that uncertainty, via the influx of adjacent electrons at this particular point, may lead to a gain in electron detection. We may calculate this gain from the normal distribution for the distance from adjacent pixels and the ratio of the diameter of a pixel, or 20pm and the circumference of a circle at this distance, so that the normalized gain $g$ from a point at distance $d$ will be (units in pm):
\begin{equation}
g(d) = \frac{20}{2 \pi d} \frac{1}{\sqrt{2 \pi \sigma^2}} \exp -\frac{d^2}{2 \sigma^2}
\end{equation}
Summing up all contributions from pixels in a distance d of less than 280 pm, we obtain a value of 0.65. For a contrast of 100, corresponding to an apparent height difference of 200pm, which is easily obtained in an STM, the gain at a particular pixel should not be larger than 0.01. Comparing with the number obtained from a normal distribution, based on a standard deviation of 280pm and a pixel width of 20pm, or 0.65, we see that such a contrast cannot be obtained if the density of charge is assumed to be a statistical quantity complying with the uncertainty relations. Again, the value obtained from the statistical analysis differs by about two orders of magnitude from the value inferred from the experiments. So that from both perspectives, the loss due to the appearance of the electron at adjacent pixels, and the gain due to the appearance of the electron statistically deviating from adjacent pixels, we  arrive at the same conclusion: the precision and contrast of today's STM experiments is well beyond the requirements of the uncertainty principle. It is thus safe to conclude that the density of charge cannot be a statistical quantity.
\begin{figure}
\centering
\includegraphics[width=\columnwidth]{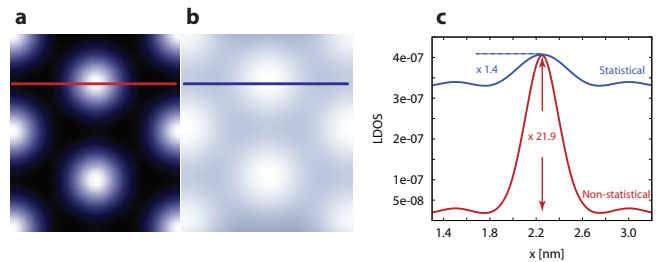}
\caption{Simulated constant height contour of silver adatom. a Simulation based on the exact values of the local density of electron charge. The contour is similar to the experimental results shown in Fig. 1 and to measurements and simulations of a Pb adatom presented in Ref. \cite{LIAN10}. b Simulation based on a convolution of probabilities for a standard deviation of 280pm. c Linescan across the adatom for both simulations. The exact values of the density yield an apparent height which is more than one order of magnitude higher than the apparent height in the simulation based on quantum statistics.}
\end{figure}
To show in detail, how a statistical density of electron charge complying with the uncertainty relations would affect the image of a single adatom at the silver surface, we have simulated the image of an adatom at very low bias voltage (20mV), in constant height mode at a distance of 500pm above the centre of the adatom. Figure 3 shows two possible images: Figure 3a reflects a non-statistical measurement, where the measured current is exactly proportional to the calculated local density of states (LDOS), this image is very close to experimental results (see Fig. 1a, and the measurements by Lian et al. \cite{LIAN10}). Figure 3b shows a hypothetical measurement, where the LDOS value at every single point is affected by the random distribution of the electrons present at a given point. In this image we have summed up all contributions from the normal distributions of adjacent measurement points, where the amplitude is equal to the exact density of states at a given point, and its contribution to adjacent points is calculated from a normal distribution with a standard deviation of 280pm. We have rescaled the ensuing LDOS values to the maximum value for the density. Figure 3c shows a simulated linescan across a single adatom. Here, the numerical difference is more than one order of magnitude. The simulation allows the conclusion that measurements of adatoms would not be able to show an apparent height difference of more than about 10pm. This is, again, a clear contradiction with experimental results, where the apparent heights are typically larger than 100pm.

In summary, we have shown that modern STM experiments violate the Heisenberg uncertainty relations by about two orders of magnitude. This indicates that the density of electron charge is not a statistical quantity, as currently believed. In a wider context, it should be noted that the question of a sub-quantum theoretical framework, something which Einstein or Schr\"odinger would have called a ''hidden variable theory'', seems to be increasingly on the agenda today. One reason for it is certainly the sophistication and precision of today's experiments, as analyzed in this paper. Another, however, is
dissatisfaction with a framework which does not seem to allow us insight into fundamental processes at the atomic scale.
 Most notable in this context are attempts to account for the probabilistic framework of quantum mechanics by an analysis of a deeper level of description, see recent papers by Khrennikov on prequantum statistical theory \cite{khrennikov11}, by Gr\"ossing et al. on subquantum thermodynamics \cite{groessing11}, by Elze on emergent quantum mechanics \cite{elze09}, and by t'Hooft on deterministic quantum mechanics \cite{thooft07}. It might be that these developments indicate that our understanding of
 reality is about to undergo a {\it quantum leap} into a new direction.

\section*{Acknowledgments}
Helpful discussions with Krisztian Palotas are gratefully acknowledged. The author thanks the Royal Society London for support.


\begin{thebibliography}{99}
\bibitem{BIN82}
G. Binnig, H. Rohrer, Ch. Gerber, E. Weibel, Phys. Rev. Lett. 49, 57 (1982).
\bibitem{BIN87}
G. Binnig, H. Rohrer, Rev. Mod. Phys. 59, 615 (1987).
\bibitem{HEL94}
E. J. Heller, M. F. Crommie, C. P. Lutz, D. M. Eigler, Nature 394, 464 (1994).
\bibitem{GAW08}
H. Gawronski, M. Mehlhorn, K. Morgenstern, K. , Science 319, 930 (2008).
\bibitem{NEE07}
N. Neel et al., Phys. Rev. Lett. 98, 016801 (2007).
\bibitem{HAR08}
K. R. Harikumar et al., Nature Nanotechnology 3, 222 (2008).
\bibitem{HOF03}
W. A. Hofer, A. S. Foster, A. L. Shluger, Rev. Mod. Phys. 75, 1287 (2003).
\bibitem{DENG06}
Z. T. Deng, et al., Phys. Rev. Lett. 96, 156102 (2006).
\bibitem{TH85}
J. Tersoff, D. R. Hamann, Phys. Rev. B 31, 805 (1985).
\bibitem{RIE04}
K.-H. Rieder et al., Phil. Trans. R. Soc. Lond. A 362, 1207 (2004).
\bibitem{HK64}
P. Hohenberg, W. Kohn, Phys. Rev. 136, B864 (1964).
\bibitem{KS65}
W. Kohn, L. J. Sham, Phys. Rev. 140, A1133 (1965).
\bibitem{BORN26}
M. Born, Zeitschr. f. Phys. 37, 863 (1926); 38, 803 (1926).
\bibitem{HEI27}
W. Heisenberg, Zeitschr. f. Phys. 43, 172 (1927).
\bibitem{HOF11}
Werner A. Hofer, Found. Phys. 41, 754 (2011).
\bibitem{SCHR26}
E. Schr\"odinger, Phys. Rev. 28,  1049 (1926).
\bibitem{HEI25}
W. Heisenberg, Zeitschr. F. Phys. 33, 879 (1925).
\bibitem{MARG68}
James L. Park and Henry Margenau, Intl. J. Theor. Phys. 3, 211 (1968). See in particular
page 213: ''Many gedankenexperiments have been designed to illustrate Heisenberg's famous law;
unfortunately, the false impression is often conveyed that his principle, which is actually
a theorem about standard deviations in collectives of measurement results, imposes restrictions
on {\it measurability}.'' [italics in the original text]
\bibitem{HEI04}
A. J. Heinrich, J. A. Gupta, C. P. Lutz, D. M. Eigler, Science 306, 466 (2004).
\bibitem{BOH52}
David Bohm, Phys. Rev. 85, 166 (1952); 85, 180 (1952).
\bibitem{DEB60}
L. De Broglie, Non-linear Wave Mechanics: A Causal Interpretation. Elsevier, Amsterdam (1960).
\bibitem{HES10}
David Hestenes, Found. Phys. 40, 1 (2010).
\bibitem{LIAN10}
J. C. Lian, et al. Phys. Rev. B 81, 195411 (2010).
\bibitem{khrennikov11}
A. Khrennikov, Europhys. Lett. 90, 40004 (2009).
\bibitem{groessing11}
G. Groessing, S. Fussy, J. Mesa Pascasio, H. Schwabl, J. Phys.: Conf. Ser. 306, 012040 (2011).
\bibitem{elze09}
H.-T. Elze, J. Phys.: Conf. Ser. 171, 012034 (2009).
\bibitem{thooft07}
G. t Hooft, J. Phys.: Conf. Ser. 67, U166 (2007).
\end{thebibliography}
\end{document}